\documentclass[aps,twocolumn]{revtex4}
\usepackage{amssymb}
\usepackage{amsmath}
\usepackage{epsfig}
\begin{document}
%




\title{Boundary instability of a two-dimensional electron fluid}                           

\author{M.I. Dyakonov}  

\affiliation{Laboratoire de Physique Th\'eorique et Astroparticules, Universit\'e 
         Montpellier 2, CNRS, France}



\begin{abstract}
It was shown previously that the current-carrying state of a Field Effect
Transistor with asymmetric source and drain boundary conditions may become
unstable against spontaneous generation of plasma waves \cite{dyakonov1}. 
By extending the analysis to the two-dimensional case we find that the dominant 
instability modes correspond to waves propagating in the direction perpendicular to 
the current and localized near the boundaries. This new type of instability should 
result in plasma turbulency with a broad frequency spectrum. More generally, it is 
shown that a similar instability might exist, when a strong enough current goes 
through a single boundary between the gated and ungated regions.


\end{abstract}

\maketitle


It was shown previously \cite{dyakonov1} that the current-carrying state of a Field
Effect Transistor may become unstable against spontaneous generation of plasma
waves in the transistor channel, provided there is an asymmetry in the boundary
conditions at the source and at the drain. An extreme case of such asymmetry is the
ac short-curcuit condition at the source and the ac open curcuit at the drain.
For submicron gate lengths the frequencies of the plasma oscillations belong to the
terahertz range, thus the FET can, in principle, serve as a generator of terahertz
radiation. The nonlinear properties of the electron fluid in the transistor channel
can be also used for detection and frequency mixing in the terahertz domain
\cite{dyakonov2}.

Experimentally, both terahertz emission \cite{knap1, knap2, knap3} and detection 
\cite{knap4} in nanometric transistors were demonstrated. Fig. 1 presents experimental 
data \cite{knap5} for a GaAlN/GaN HEMT at 4 K clearly showing the emission threshold at 
a certain source-drain voltage (or current) and a typical broad emission spectrum in the 
terahertz domain. Contrary to the prediction of Ref. \cite{dyakonov1}, the spectrum  
depends neither on the gate length, nor on the gate voltage. Similar results for terahertz 
emission were obtained at room temperature \cite{knap3}.

It is not firmly established that the observed emission is indeed related to the instability
predicted in \cite{dyakonov1} (see \cite{knap3}). However, one cannot directly compare
the theory with the experiments because the experimental geometry is very
different from the one-dimensional model adopted in \cite{dyakonov1}. In the standard
experimental situation, the width of the gate $W$ is much larger than the gate length $L$, 
typically $W/L\sim 100$, see Fig. 2, left. Under such conditions the
one-dimensional model, where the plasma density and velocity depend on the coordinate $x$ 
only, is not appropriate, since obviously oblique plasma waves with a non-zero component of 
the wave vector in the $y$ direction can propagate. In such a geometry, the gated region is 
not a resonator, but rather a waveguide with a continuous spectrum of plasma waves, see Fig. 2,
right.

The purpose of this work is to extend the analysis of stability of the steady-state flow 
\cite{dyakonov1} to the more realistic geometry of Fig. 2. Since $W>>L$, we will consider 
the limit of a strip, which is infinite in the $y$ direction. It will be demonstrated that 
in such a geometry a new mode of instability dominates, which is localized near the gate 
boundaries. Moreover, a similar instability should exist in the limit $L\rightarrow \infty$, 
i.e. near a single boundary of a current-carrying two-dimensional plasma. 

\begin{figure}
\epsfxsize=240pt {\epsffile{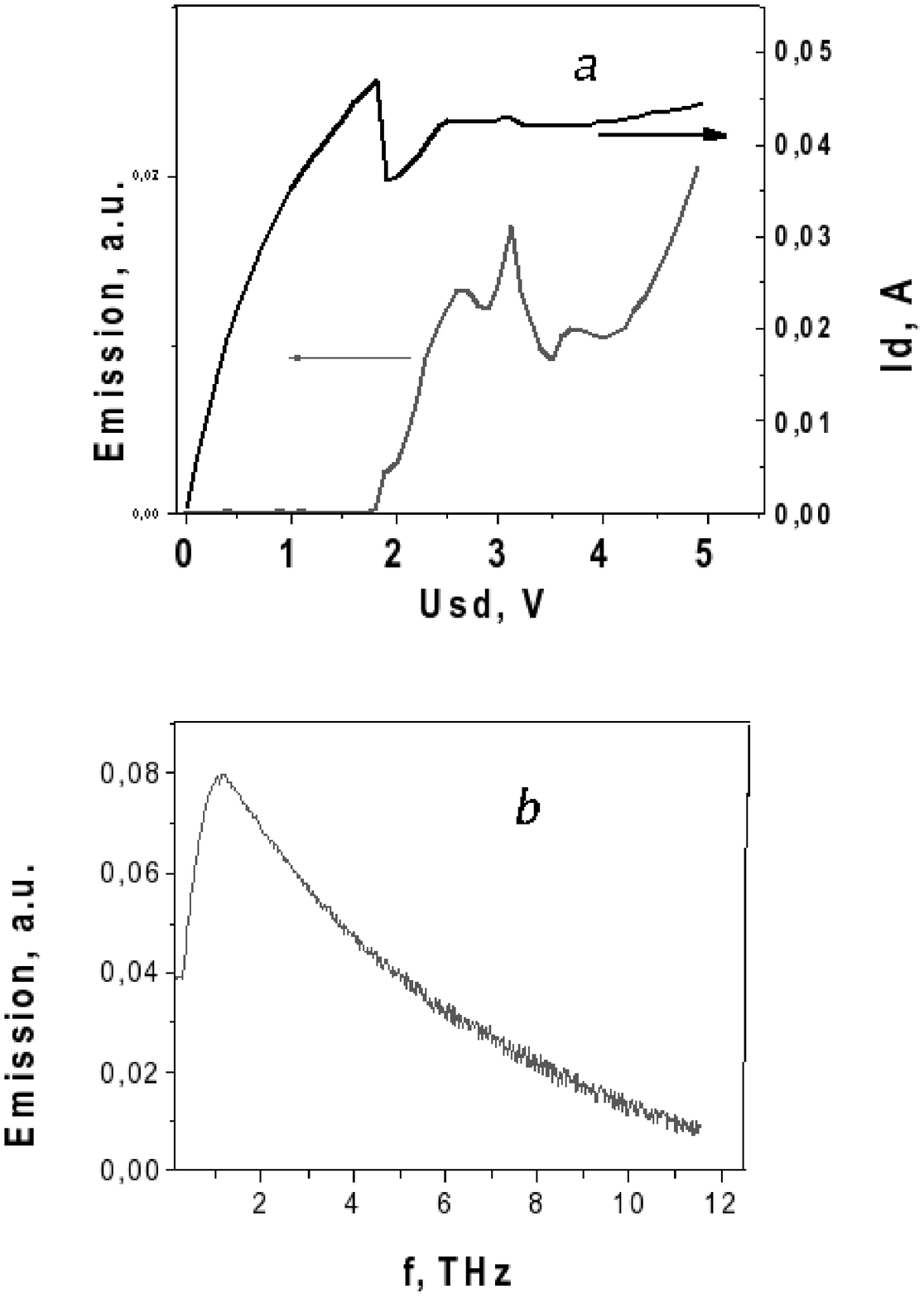}}
\caption{Experimental results for THz emission from a AlGaN/GaN HEMT at 4.2 K \cite{knap5}.
\textit{a} -- The drain current (right scale) and the emission intensity (left scale), as 
functions of the source-drain voltage, $U_{sd}$. Note the pronounced threshold for 
emission. \textit{b} -- The emission spectrum at $U_{sd}= 3$\,V. }
\end{figure}

\begin{figure}
\epsfxsize=240pt {\epsffile{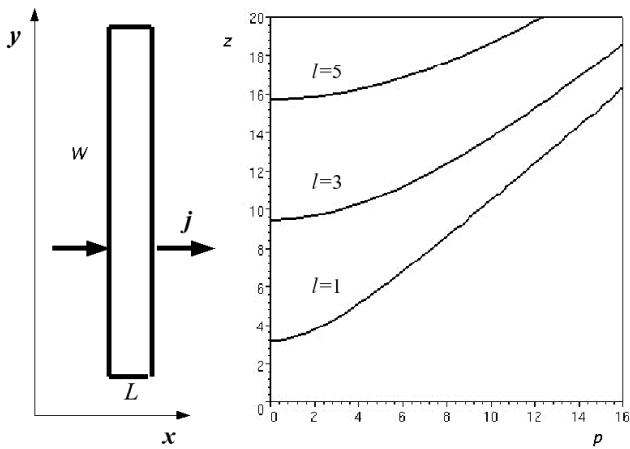}}
\caption{Left panel: the geometry of the gate. The width $W$ is much greater than the length 
$L$. Right panel: The plasma wave spectrum in a strip, $z=2L\omega/s$ is the dimensionless
frequency, $p=2qL/s$ is the dimensionless wave vector in the $y$ direction, $s$ is the plasma
wave velocity.}
\end{figure}
Within the hydrodynamic approach the electrons in a gated 2D channel can be 
described by the following equations \cite{dyakonov1}:
$$ \frac{\partial {\bf V}} {\partial t}+({\bf V} {\bf \nabla}){\bf V} = -\frac{e} {m} 
\nabla U, \eqno{(1)}$$
$$ \frac{\partial U}{\partial t}+ {\bf \nabla }(U {\bf V})=0, \eqno{(2)}$$
where ${\bf V}({\bf r},t)$ is the electron hydrodynamic velocity,  $U({\bf r},t)$
is the gate-to-channel voltage swing, ${\bf r}$ is the vector in the 2D plane, $e$
and $m$ are the electron charge and the effective mass respectively. Equation (1)
is the Euler equation, and Eq.(2) is, in fact, the continuity equation since the
electron density in the channel, $n$, is related to the voltage swing, $U$, by the
relation
$$ en= CU, \eqno{(3)}$$
where $C$ is the gate to channel capacitance per unit area. This equation holds if
the scale of the variation of the potential in the channel is large compared to the
gate-to-channel separation $d$ (the graduate channel approximation).

Collision processes give an additional term $-{\bf V}/\tau$ in the right-hand side
of Eq.(1), where $\tau $ is the momentum relaxation time. In the following, this term 
will be neglected, however it should be understood that the instabilities studied below 
will practically exist only if the instability increment is greater than $1/\tau$, a 
condition that determines the instability threshold for the drift velocity, similar to 
the situation in the one-dimensinal model \cite{dyakonov1}.

We chose the $x$ axis in the direction from source ($x=0$) to drain ($x=L$) and, 
following Ref. 1, we impose the asymmetric boundary conditions of a fixed voltage
at the source and a fixed current at the drain: $ U=U_0$ at $x=0$ and $j_x=j_0$ 
at $x=L$, where $j_x$ is the $x$ component of the current density. Because
of Eq. (3), the latter condition can be rewritten as $(UV_x)_{x=L}=U_0 v_0$, where
$v_0=j_0/en$ is the electron drift velocity.

It was pointed out in \cite{dyakonov1} that Eqs. (1) and (2) are identical to the 
equations
describing the so called "shallow water" in conventional hydrodynamics 
\cite{landau}, plasma waves in the channel being analogous to water
waves in the case when the wavelength is much larger than the water depth.
Furthermore, it was shown that the current-carrying steady state described by the 
stationary solution of Eqs. (1,2) with the above boundary conditions, $U=U_0, 
V_x=v_0$, is unstable against spontaneous generation of plasma waves with a growth
increment given by
$$ \gamma = \frac{s^2-v_{0}^2}{2sL}\ln \bigl |\frac{s+v_0 }{s-v_0} \bigr|, \eqno{(4)}$$
where $s=(eU_0/m)^{1/2}$ is the plasma wave velocity.

This result followed from a one-dimensional analysis, e.g. small perturbations of
the steady state were assumed to be independent of the coordinate $y$ in the
direction perpendicular to the current. As we shall see, the extension of the
analysis to $y$-dependent perturbations not only gives corrections to Eq. (4) but,
somewhat unexpectedly, gives a new mode of instability which always dominates.

We study the time dependence of small perturbations of the steady state.
Accordingly we put $U=U_0 + (m/e)u, V_x = v_0 +v_x, V_y=v_y$ and we linearize
Eqs. (1,2) with respect to the small quantities $u, v_x,v_y$. 

The boundary conditions become: 
$$u_{x=0}=0, \hspace{0.2in} (v_0u+s^2 v_x)_{x=L}=0, \eqno{(5)}$$ 
(zero ac voltage at the source and zero ac current at the drain). We look for the 
solutions of the linearized equations with $u, v_x, v_y \sim \exp (-i\omega t+ikx+iqy
)$, where $k$ and $q$ are the components of the wave vector in the $x$ and $y$ 
directions respectively. This procedure gives
$$(\omega -kv_0)v_x-ku=0, \eqno{(6)}$$
$$(\omega -kv_0)v_y-qu=0, \eqno{(7)}$$ 
$$(\omega -kv_0)u-s^2(kv_x+qv_y)=0.  \eqno{(8)}$$
The dispersion relation for the plasma waves follows:
$$ (\omega -kv_0)^2=s^2(k^2+q^2),\eqno{(9)}$$
the term $kv_0$ taking into account the Doppler shift due to the motion of the
electron fluid. For given $\omega$ and $q$ we find two values for the $x$-component
of the wave vector, corresponding to oblique waves propagating downstream and 
upstream:
$$k_{1,2}= \frac{-\omega v_0 \pm s\sqrt {\omega^2 -(s^2-v_0 ^2)q^2}}{s^2-v_0 ^2}.
\eqno{(10)}$$
For $q=0$ this reduces to $k_1= \omega /(s+v_0), k_2 =-\omega /(s-v_0)$. The
general solution for $u$ and $v_x$ can be found using Eq. (6):
$$u = A\exp (ik_1 x)+B\exp (ik_2 x),\eqno{(11)}$$
$$v_x = \frac{k_1}{\omega -k_1 v_0}A\exp (ik_1 x)+\frac{k_2}
{\omega -k_2 v_0}B\exp (ik_2 x),\eqno{(12)}$$
where $A$ and $B$ are constants.

The boundary conditions, Eq. (5), together with Eq. (10) give the relation
$$\exp \bigl (i(k_1 -k_2)L \bigr )= -\frac{\omega -k_1 v_0}{\omega -k_2 v_0}, 
\eqno{(13)}$$
which can be rewritten in the form:
$$\exp (i\sqrt {z^2-p^2})= \frac{\beta \sqrt {z^2-p^2}-z}{\beta \sqrt {z^2-p^2}+z},
\eqno{(14)}$$
where the dimensionless variables for frequency, wave vector, and drift velocity
are introduced:
$$z=\frac{2sL}{ s^2-v_0^2} \omega, \hspace{0.2in} p=\frac{2sL}{\sqrt{s^2-v_0^2}}q,
\hspace{0.2in} \beta =\frac{v_0}{s}. \eqno{(15)}$$

Equations (14), (15) define the complex frequency $\omega = \omega' +i\gamma$ as a
function of the drift velocity $v_0$ and the wave vector $q$. For $q=0$ one obtains
the previous one-dimensional result \cite{dyakonov1} with $\omega' =\pi l(s^2-v_0^2)
/(2sL)$, where $l$ is an odd integer, and the increment $\omega ''=\gamma$ given 
by Eq. (4). 

In the general case Eq. (14) can be solved only numerically. However, an analytical 
solution can be obtained for drift velocities small compared to the plasma wave 
velocity ($\beta << 1$). For $\beta =0$ the solution of Eq. (14) is $z=(l^2+p^2)^{1/2}$, 
or in dimensional units $\omega =s\bigl ( (\pi l/L)^2+q^2\bigr )
^{1/2}$, which represents the spectrum of plasma waves in an infinite
strip with the assumed boundary conditions at $x=0$ and $x=L$ (Fig. 2). The linear in
$\beta$ correction to this value is purely imaginary:
$$\gamma = \frac{v_0}{L}\frac{1}{1+(qL/(\pi l))^2}. \eqno{(16)}$$

Thus, as $q$ increases and becomes comparable to or larger than the quantized value
of $k=\pi l/L$ for the $l$-th mode, the instability increment decreases from its 
value $v_0/L$ given by Eq. (4) for $v_0<<s$. The correction to the real part of 
$\omega$ is of second order in $\beta$.

However, in addition to this predictable result, another solution of Eq. (14) 
exists, for which $z$ (or $\omega$) is purely imaginary. For $v_0<<s$ this
solution can be found analytically by assuming that $|z|<<p$ and 
$(z^2-p^2)^{1/2} \approx ip$. This gives $z=i\beta p\tanh(p/2)$ or, in dimensional
units $\omega'=0$ and
$$\gamma = qv_0\tanh (qL). \eqno{(17)}$$
For large $qL$ this gives $\gamma =|q|v_0$, thus in contrast to the result given
by Eq. (16), the growth increment of this new mode increases at large $q$, so that
this mode of instability is the dominant one. The numerical solutions of Eq. (14) for 
$\beta =0.5$ are presented in the Fig. 3, together with the approximate result for
$\beta <<1$ given by Eq (17). 

\begin{figure}
\epsfxsize=200pt {\epsffile{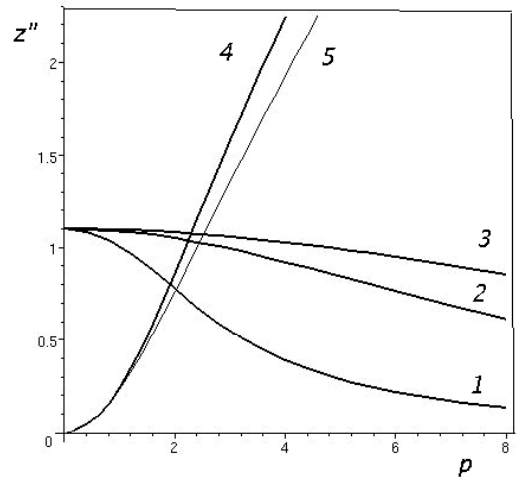}}
\caption{The instability increment $z''$ as a function of the transverse component of the 
wave vector $p$ in dimensionless units for $\beta =v_0/s=0.5$ (Eqs. (14, 15), numerical 
calculation). \textit{1, 2, 3} -- for normal modes with $l=1$, 3, and 5, respectively, 
\textit{4} -- for the new mode of instability, \textit{5} -- approximation given by Eq. (17).}
\end{figure}

It can be seen that for $qL>>1$ the new mode is localized near the boundaries at $x=0$ and 
$x=L$ on a distance $\sim 1/q$. For example, in the case $\beta <<1,qL>>1$ we have $k_1 
\approx -k_2 \approx iq$ (see Eqs. (9,10)). Thus the instability mode is formed by waves 
which are evanescent in the $x$ direction.

Since for large $|q|L$ the growth increment for the new mode does not depend on $L$, and since 
in this case the mode is localized near the boundaries, it seems plausible that a similar
instability of the steady-state flow should exist for a \textit{single} boundary of an infinite
(both in the $y$ and the $x$ directions) two dimensional current-carrying plasma. We now show 
that this is indeed the case.

Let a steady current with the drift velocity $v_0$ flow across the boundary ($x=0$) of a 
semi-infinite sample situated at $x>0$. The general boundary condition at $x=0$ is defined by 
the impedance $\zeta$ relating the ac voltage and the ac current (compare with Eq. (5)):
$$u=\zeta (v_0u+s^2v_x). \eqno{(18)}$$
The boundary condition at $x=\infty$ corresponds to the vanishing of the small perturbations, 
$u= v_x =v_y=0$. 

The impedance $\zeta$ will be considered as purely imaginary: $\zeta =i\lambda/s$, where 
$\lambda$ is the dimensionless parameter proportional to the effective capacitance. (The 
existence of a finite resistance, described by the real part of $\zeta$, will obviously 
introduce damping of the initial perturbations and, if it is large enough, any instability 
will be supressed).
 
To insure the boundary condition at $x=\infty$, we now keep only one exponent in Eqs. (11,12),
with the wave vector $k$, whose imaginary part is positive. These equations, together with Eq. 
(18), give:
$$\omega -kv_0= \alpha sk, \qquad \alpha =\frac{i\lambda}{1-i\lambda \beta}. \eqno{(19)}$$
Inserting this relation in Eq. (9), we find the value of the wave vector 
$k=i|q|(1-\alpha ^2)^{-1/2}$, where the sign of the square root should be chosen so that its 
real part be positive. 

Finally, from Eq. (19) one finds $\omega$. Its imaginary part $\gamma$ defines the instability 
increment:
$$\gamma = G|q|v_0, \qquad 
G=\frac{1}{\beta}\text{Re} \Bigl (\frac {\alpha +\beta}{\sqrt{1-\alpha ^2}}\Bigr ). \eqno(20)$$

Note, that the values of $q$ in Eqs. (17,20) are limited by the condition $q<1/d$, 
where $d$ is the gate-to-channel separation. For larger $q$ the graduate channel approximation,
used in deriving Eqs. (1,2) breaks down.

\begin{figure}
\epsfxsize=200pt {\epsffile{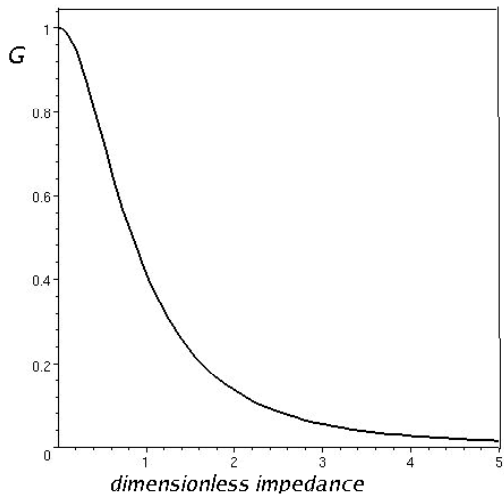}}
\caption{The coefficient $G$ in Eq. (20) for $\beta=v_0/s=0.5$ as a function of the 
dimensionless parameter $\lambda$ defining the boundary condition at $x=0$ (the boundary 
impedance is presented as $\zeta = i\lambda/s$). For large $\lambda$, $G\sim 1/\lambda^3$.}
\end{figure}

The dimensionless coefficient $G$ depends on the value of $\lambda$, defining the boundary 
impedance, and on the flow velocity $v_0$, see Eq. (19). For $\lambda<<1$, we have $G=1$ and
Eq. (20) coincides with Eq. (17) for large $|q|L$. With increasing $\lambda$ the coefficient 
$G$ decreases (see Fig. 4), reducing the instability increment, which however remains always
positive.

Thus, if the condition $|q|v_0 > 1/\tau$ is satisfied, the current-carrying steady state is
unstable against small perturbations, and the region of instability is localized near the
boundary. This is similar to what one observes in a river, when the water flows with 
sufficient velocity across an abrupt step in the waterbed: waves with wave vectors 
perpendicular to the flow are excited, while the wave vectors in the direction of the flow 
are purely imaginary, which accounts for the localization of the turbulent region near the 
step.

Certainly, the linear theory cannot predict the outcome of this instability. However,
since the spectrum of plasma waves is continuous, it seems likely that the instability will 
result in a turbulent motion of the electron fluid near the boundary of the gated region. The 
spectrum of the plasma oscillations should be broad, similar to what is observed in 
experiments (Fig. 1). The width of the spectrum is expected to be limited by the value
$\omega_{max} \sim s/d$, where $d$ is the gate-to-channel separation (see above). 

The present theory can be also applied to the ungated electron fluid (analogous to the "deep 
water" in conventional hydrodynamics). It was shown \cite{dyakonov3} that a one-dimensional 
instability similar to the one described in Ref. \cite{dyakonov1} should exist in the ungated 
case too, under appropriate boundary conditions. It can be easily shown, that the boundary 
instability considered here will also occur in the ungated region, if the drift velocity is 
directed \textit{inside} this region, similar to the results given by Eqs. (17,20) for the 
gated electron fluid. Thus, at the boundary between the gated and ungated regions, the 
turbulence should always appear on the downstream side.

It appears that the above concept acounts for the most important experimental observations 
\cite{knap1, knap2, knap3}: the sharp threshold for terahertz emission and the broad emission
spectrum, which does not depend on the gate length, and only weakly depends on the gate potential.
A possible check of the proposed explanation would be to isolate the emission coming from one 
gate edge and to verify that the emission intensity (and possibly its spectrum) depends on the 
direction of the the drift velocity. 

On the theoretical side, the very difficult issue of the true conditions at the boundary
between the ungated and gated regions should be elucidated. (From the hydrodynamical point of
view this is the problem of what happens for a flow across the boundary between deep 
and shallow water). Also, the role of the viscosity of the electron fluid, which may supress the 
instability for large wave vectors $q$, remains to be understood.

I thank Wojciech Knap, Nina Dyakonova, Michael Shur, and Maria Lifshits for numerous helpful
discussions.


\end{document}